\documentclass[ prd, twocolumn, nofootinbib, floatfix]{revtex4-1} 
\usepackage{mathptmx}
\DeclareMathAlphabet{\mathcal}{OMS}{cmsy}{m}{n}

\usepackage[mathscr]{euscript}
\usepackage{amsmath}
\usepackage{graphicx}
\usepackage{dcolumn}
\usepackage{bm}
\usepackage{epsfig}
\usepackage{amssymb,latexsym,mathrsfs}
\usepackage{graphicx}
\usepackage{color}
\usepackage{hyperref}
\usepackage{float}
\usepackage{diagbox}
\definecolor{vividviolet}{rgb}{0.62, 0.0, 1.0}
\definecolor{amaranth}{rgb}{0.9, 0.17, 0.31}
\definecolor{palatinateblue}{rgb}{0.15, 0.23, 0.89}
\definecolor{brightpink}{rgb}{1.0, 0.0, 0.5}
\definecolor{cornflowerblue}{rgb}{0.39, 0.58, 0.93}
\definecolor{deepcarminepink}{rgb}{0.94, 0.19, 0.22}
\definecolor{radicalred}{rgb}{1.0, 0.21, 0.37}
\definecolor{blueblue}{RGB}{21,47,181}
\definecolor{greengreen}{RGB}{65,166,16}

\hypersetup{ linktoc=all,
    colorlinks, linkcolor={blueblue},
    citecolor={greengreen}, urlcolor={blueblue}
}

\newcommand{\be}{\begin{equation}}
\newcommand{\ee}{\end{equation}}
\newcommand{\bs}{\begin{split}} 
\newcommand{\bea}{\begin{eqnarray}}
\newcommand{\eea}{\end{eqnarray}}

\newcommand{\D}{\mathrm{d}}
\newsavebox{\myhbar}

\begin{document}

%\title{(1+1)-dimensional Hawking radiation model for a Kerr-Newman black hole} 
\title{Hawking radiation particle spectrum of a Kerr-Newman black hole}

\author{Joshua Foo${}^{1}$}
\email{joshua.foo@uqconnect.edu.au}
\author{Michael R.R. Good${}^{2}$}
\email{michael.good@nu.edu.kz}
\affiliation{${}^1$Centre for Quantum Computation \& Communication Technology, School of Mathematics \& Physics,
The University of Queensland, St.~Lucia, Queensland, 4072, Australia\\
${}^2$Physics Department \& Energetic Cosmos Laboratory, Nazarbayev University, 
Nur-Sultan, Kazakhstan\\
%${}^3$Energetic Cosmos Laboratory, Nazarbayev University, Nur-Sultan, Kazakhstan\\ 
}

\begin{abstract} 
Charged, rotating Kerr-Newman black holes represent the most general class of asymptotically flat black hole solutions to the Einstein-Maxwell equations of general relativity. Here, we consider a simplified model for the Hawking radiation produced by a Kerr-Newman black hole by utilising a (1+1)-dimensional accelerated boundary correspondence (i.e.\ a flat spacetime mirror trajectory) in Minkowski spacetime. We derive the particle spectrum of the outgoing massless, scalar field and its late-time thermal distribution which reduces to the Kerr, Reissner-Nordstr\"om and Schwarzschild cases in the appropriate limits. We also compute the particle spectrum of the extremal Kerr-Newman system, showing that the total energy emitted is finite.  
\end{abstract} 

\date{\today} 

\maketitle 

\section{Introduction}\label{sec:Intro}
According to the no-hair theorem, black hole solutions to the Einstein-Maxwell equations of general relativity -- combining the field equations of gravity and electromagnetism -- are fully characterised by their mass, $M$, angular momentum, $J$ and electric charge, $Q$. The Kerr-Newman solution \cite{Newman:1965my} for a charged, rotating black hole represents the most general class of asymptotically flat black holes. As such, it continues to be of significant theoretical and mathematical interest, providing an entry point for further exploration into the numerous electrovacuum solutions \cite{Adamo:2014baa} to the Einstein–Maxwell equations. 

In this paper, we derive a simplified (1+1)-dimensional model of Hawking radiation \cite{Hawking:1974sw, Davies:1976hi,Davies:1977yv} emitted by a Kerr-Newman black hole, known as an accelerated boundary correspondence (ABC). It is well-known that accelerated boundaries (i.e.\ perfectly reflecting mirrors) in flat spacetime radiate particles \cite{moore1970quantum,DeWitt:1975ys}, induced by the rapid changing of boundary conditions on incoming vacuum modes. For appropriately chosen trajectories  \cite{Good:2018aer,Good:2017ddq,Good:2017kjr}, close comparisons can be made with the radiation emitted from black holes formed via gravitational collapse (neglecting greybody factors and higher-dimensional effects) \cite{wilczek1993quantum}. While the correspondence has been well-studied in the literature (e.g.\ Calogeracos asymptotics \cite{Calogeracos:2001nf,Calogeracos:2001ne,Calogeracos:2002qh,Calogeracos:2004wg}, Ritus duality \cite{Ritus:2003wu,Ritus:2002rq,Ritus:1999eu,Nikishov:1995qs}, equivalence principle controversies \cite{Fulling:2018lez,fullingpage}, entanglement harvesting and entropy evolution \cite{Cong:2018vqx,Cong:2020nec,Chen:2017lum,Lee:2019adw,Good:2018ell,Bianchi:2014qua,Bianchi:2014vea}, mirror partner particles \cite{Tomitsuka:2019ehm,Hotta:2015yla}, uniformly and exponentially accelerated analogs \cite{Kay:2015iwa,Obadia:2002ch,Obadia:2002qe,Juarez-Aubry:2014jba,Foo_2020}, and plasma mirrors \cite{Chen:2015bcg,Chen:2020sir}), recent works have applied the ABC to previously unstudied spacetimes including the Kerr \cite{Good:2020fjz}, Reissner-Nordstr\"om (RN) \cite{good2020particle}, de Sitter and anti-de Sitter geometries \cite{Good:2020byh}, extremal RN \cite{good2020extreme,Liberati:2000sq} and Kerr \cite{Good:2020fjz,Rothman:2000mm}, showing its continued relevance \cite{Lin:2020itp,Good:2020uff,Alves:2020vgl} as an analytical tool in curved spacetime quantum field theory \cite{Fabbri, Birrell:1982ix}. The main novelty of these recent works has been in demonstrating that the intrinsic properties of the black hole -- including its mass, charge and angular momentum -- can be encoded within the noninertial motion of the accelerated mirror. In this paper, we offer an extension of the mirror model to include \textit{non-spherically symmetric} (Kerr-Newman) black holes, showing that it correctly predicts the Hawking temperature and spectral distribution of the radiation. This result is surprising, since the correspondence between the (1+1)- and (3+1)-dimensional systems is derived by flattening out two spatial dimensions described by the angular and azimuthal coordinates $\theta$ and $\phi$, which are key in the description of the black hole's angular momentum. Our results demonstrate the ongoing explanatory power and calculational simplicity of the straightforward yet highly instructive model, as well as providing clear insights about the interplay between the mass, charge and angular momentum and their effect on the outgoing particle and energy spectrum. 

Our paper is organized as follows: in Sec.\ \ref{sec:metacc}, we review the Kerr-Newman metric and the transformation of this coordinate system to an accelerated mirror trajectory. In Sec.\ \ref{sec:particles}, we derive the energy and particle flux of analog Hawking radiation emitted by the mirror. The thermal character of this radiation is demonstrated for late-times, as well as its novel dependence on the charge and angular momentum parameters of the black hole system. In Sec.\ \ref{sec:extremal}, we extend our study to the extremal Kerr-Newman limit and study the radiative properties of the corresponding mirror trajectory. Throughout, we make comparisons to the Kerr, RN and Schwarzschild metrics, confirming prior results \cite{Good:2016oey,Good_2017Reflections,Anderson_2017,Good_2017BHII}. In this paper, we adopt natural units, $G = \hslash = c = k_B = 1$.

%%%%%%%%%%%%%%%%%%%%%%%%%%%%%% 
\section{Kerr-Newman Black Holes} \label{sec:metacc} 
\noindent The Kerr-Newman (KN) metric takes the form,
\begin{align}\label{KN}
    \D s^2 &= - \frac{\Delta}{\rho^2} \big( \D t - a \sin^2\theta \D \phi \big)^2 + \frac{\rho^2}{\Delta} \big( \D r^2 + \Delta \D \theta^2 \big) \nonumber \\
    & + \frac{\sin^2\theta}{\rho^2 } \big( ( r^2 + a^2 ) \D \phi - a \D t \big)^2,
\end{align}
where $a= J/M$ is the mass-normalised angular momentum, $\rho^2 = r^2 + a^2 \cos^2\theta \vphantom{\frac{J}{M}}$ and $\Delta = r^2 - r_S r + a^2 + Q^2 \vphantom{\frac{J}{M}}$.  Here $r_S = 2M$ is the usual Schwarzschild radius, and $Q$ is the electric charge of the black hole. Following \cite{Good:2020fjz}, let us restrict our analysis to a plane of constant $\theta, \phi$
which yields the simplified (1+1)-dimensional metric
\begin{align}
    \D s^2 &= - f(r) \D t^2 + \frac{\D r^2}{f(r)},
\end{align}
where
\begin{align}\label{28}
    f(r) &= \frac{r^2 - 2Mr  + a^2 + Q^2}{r^2 + a^2\cos^2\theta} .
\end{align}
The metric function of Eq.\ (\ref{28}) is independent of $\phi$ and so the temperature of the Hawking radiation will likewise be unaffected by changes in $\phi$. However the presence of the angular coordinate $\theta$ will leave an angular dependence in the temperature itself. To understand this, our model derives a correspondence between the (3+1)-dimensional black hole coordinates and the (1+1)-dimensional flat spacetime mirror trajectory by flattening out the two additional spatial dimensions defined by $\theta$, $\phi$. However the rotational degree of freedom in the full Kerr-Newman metric, Eq. (\ref{KN}), creates an ergosphere outside the black hole defined by $r_+ < r < r_{e+}$ where
\begin{align}\label{29}
    r_{e+} &= M + \sqrt{M^2  - Q^2 - a^2 \cos^2\theta}.
\end{align}
We find that $r_{e+} = r_+$ (the outer horizon) when $\theta = 0$. In deriving the accelerated boundary correspondence between the black hole and the flat spacetime mirror trajectory, we require a tortoise coordinate defined with respect to the outer horizon, $r_+$, of the black hole. To obtain a physically meaningful tortoise coordinate, we therefore require $\theta = 0$ which should (and as we show, does) yield the correct outer horizon surface gravity and likewise, the temperature. If $\theta \neq 0$, then one has an ill-defined tortoise coordinate which does not actually correspond to radial coordinate of the outer horizon. 

In light of this, we set $\theta = 0$, $\phi = \text{const.}$ in Eq.\ (\ref{KN}) which yields the corresponding (1+1)-dimensional line element, 
\begin{align}\label{KN1}
    \D s^2 &= - f(r) \D t^2 + f(r) \D r^2, 
\end{align}
where
\begin{align}
    f(r) &= 1 - \frac{2Mr}{r^2 + a^2} + \frac{Q^2}{r^2 + a^2} \label{f(r)}. 
\end{align}
It is clear that in the limit $Q\to 0$, Eq.\ (\ref{KN1}) reduces to the uncharged, (1+1)-dimensional Kerr metric \cite{Good:2020fjz}; conversely, as $a\to 0$, one obtains the (1+1)-dimensional RN spacetime \cite{good2020particle}. Notice that Eq.\ (\ref{KN1}) possesses two horizons at the radial coordinates $r_\pm = M \pm \sqrt{M^2 - a^2 - Q^2}$ which reduce to the event horizon, $r = r_S$, and the coordinate singularity $r = 0$ of the Schwarzschild black hole, in the limit $a , Q \to 0$. We have introduced this (1+1)-dimensional metric to find the associated radial trajectory (i.e.\ the mapping of the interior coordinates to the exterior coordinates, see Sec. \ref{sec:K}) of the center of the black hole in (3+1) dimensions. As noted in previous works \cite{wilczek1993quantum, Good:2020fjz, good2020particle,Good:2020byh,good2020extreme}, this acts as the reflecting point of incoming vacuum modes. In the accelerated boundary correspondence, the trajectory of the mirror functions as this reflecting point in flat, Minkowski spacetime. Since the angular coordinates $\theta$ and $\phi$ do not enter the line element in Eq.\ (\ref{KN1}), the radiation flux of the outgoing modes will be not depend on these parameters. The temperature of outgoing radiation will be identical to that obtained for the (3+1)-dimensional case, modulo greybody factors and higher-dimensional effects, as we discuss in Sec.\ \ref{sec:particles}.

%%%%%%%%%%%%%%%%%%%%%% 
\subsection{(1+1)-dimensional mirror trajectory} \label{sec:K} 
In the Kerr-Newman spacetime, the thermal radiation emitted from the black hole and detected by an inertial observer at infinity has temperature
\begin{align}
    T_\text{KN} &= \frac{\kappa_+}{2\pi} = \frac{1}{4\pi} \frac{r_+ - r_-}{r_+^2 + a^2}, \label{KNtemp}
\end{align}
where $\kappa_+$ is the surface gravity of the outer horizon. For a double null coordinate system ($u,v$) with $u = t-r^\star$ and $v = t + r^\star$, the associated tortoise coordinate $r^\star$ is obtained \cite{Fabbri},
\begin{align}
    r^\star &= \int \frac{\D r }{f(r)},\label{findtort}
\end{align}
which yields
\begin{align}\label{tortoise2}
    r^\star &= r + M \ln \bigg| \frac{(r - r_+)( r - r_-)}{r_S^2} \bigg| + \frac{2M^2 - Q^2}{2\lambda}  \ln \bigg| \frac{r - r_+}{r - r_-} \bigg|, 
\end{align}
where $\lambda = \sqrt{M^2 - a^2 - Q^2}$.
The metric for the geometry describing the outside collapsing region takes the simplified form, $\D s^2 = - f\: \D u \D v $.  The matching condition (see \cite{Fabbri}) with the flat interior geometry, described by the interior coordinates $U = T - r$ and $V = T + r$ is the trajectory corresponding to $r = 0$, expressed in terms of the exterior function $u(U)$ with interior coordinate $U$. We can obtain this matching via the association $r = r^\star$, and taking $r^\star ( r = (v_0 - U) /2 ) = (v_0 - u )/2$ along a light ray, $v_0$. In a similar manner to the uncharged Kerr black hole \cite{Good:2020fjz}, we can choose $v_0  - 2r_+ \equiv v_H$ or $v_0 - 2r_- \equiv v_H$, since $u \to \infty$ at $U = v_H$. Without loss of generality, we set $v_H= 0$ and neglect the inner horizon solution so that $v_0 = 2r_+$. This choice is justified since it reduces to the correct Schwarzschild limit, wherein $r_- = 0$ represents the curvature singularity. 

Doing so yields the following expression for the exterior coordinate, 
\begin{align}\label{KNU}
    u(U) &= U -  \frac{1}{\kappa_+} \ln \bigg| \frac{U}{4M} \bigg| - \frac{1}{\kappa_-} \ln \bigg| \frac{U - 4\lambda}{4M} \bigg|, 
\end{align}
where
\begin{align}\label{inversesurface}
    \frac{1}{\kappa_\pm } &= 2M \pm \frac{2M^2 - Q^2}{\sqrt{M^2 -a^2 -Q^2}},
\end{align}
are the inverted surface gravities at the outer and inner horizons (corresponding to $\kappa_+,\kappa_-$) respectively. It can be straightforwardly verified that Eq.\ (\ref{KNU}) reduces to the Schwarzschild ($a,Q = 0$), Kerr ($Q = 0$), and RN $(a = 0$) exterior coordinates in the appropriate limits. 

The regularity condition of the modes requires that they vanish at $r = 0$, which acts as the reflecting boundary in the black hole coordinate system. In the accelerated boundary correspondence, the origin functions as the mirror trajectory in the $(U,V)$ coordinates. Since the field vanishes for $r<0$, the form of the field modes can be determined, allowing for the identification $U\Leftrightarrow v$ in the Doppler-shifted modes. One can define the ray-tracing function, $f(v)$, for the analog Kerr-Newman mirror trajectory by making the further identification $u(U) \Leftrightarrow f(v)$, which is a known function of the advanced time $v$. 
\begin{figure}[h]
    \centering
    \includegraphics[width=0.9\linewidth]{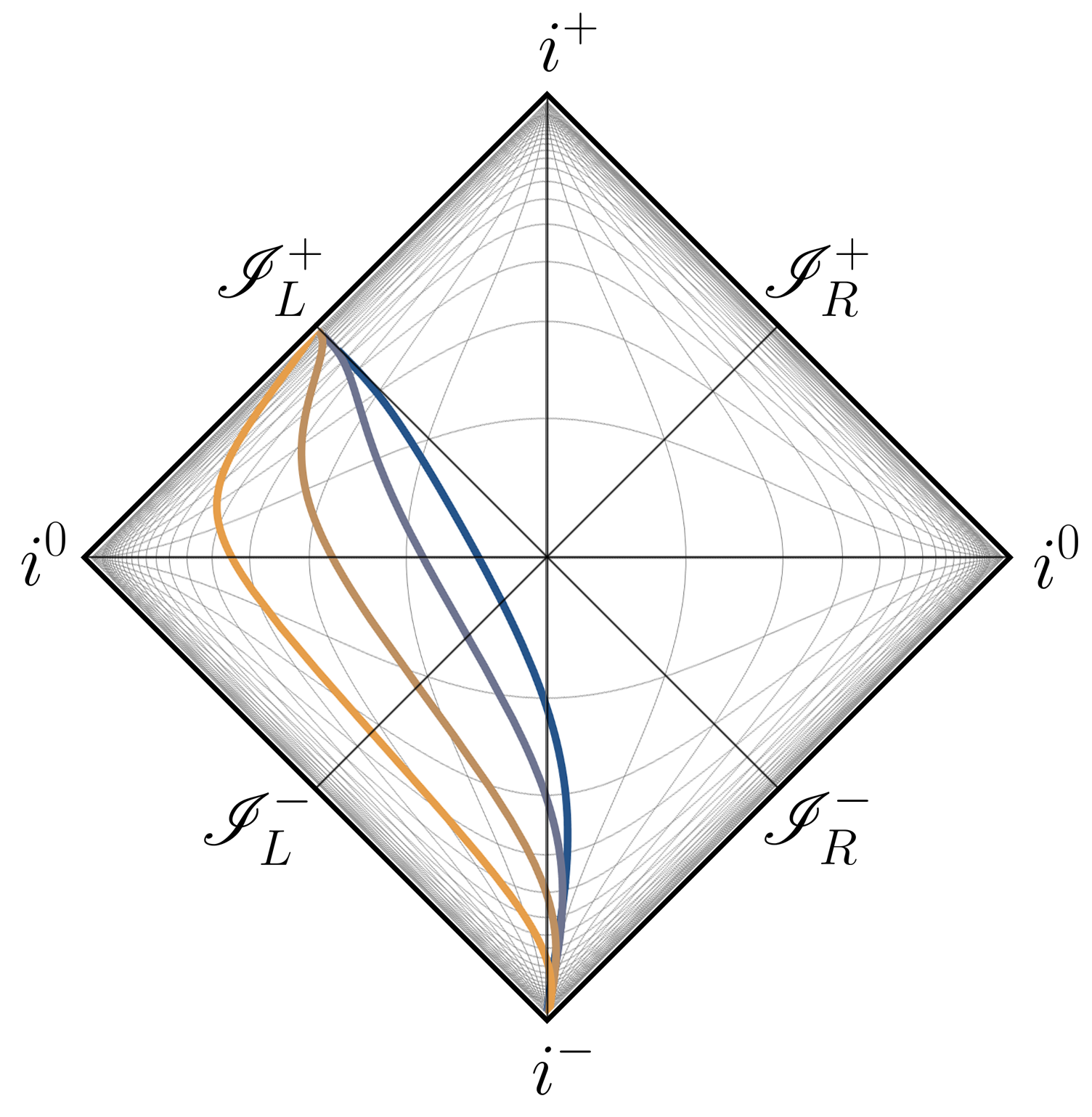}
    \caption{Penrose conformal diagram of the mirror trajectories, Eq.~(\ref{f(v)}), with $M = 1/8,1/4,1/2,1$, ranging from blue to orange respectively, and $a = Q = 1/16$. } 
    \label{fig:penrose}
\end{figure}

In the following, we consider a massless scalar field in (1+1)-dimensional Minkowski spacetime. From Eq.\ (\ref{KNU}), the ray-tracing function of the mirror is given by 
\begin{align}
    f(v) &= v - \frac{1}{\kappa_{+}}\ln | \kappa v | - \frac{1}{\kappa_-} \ln | \kappa( v - 4\lambda ) |, \label{f(v)}
\end{align}
where $\kappa = 1/4M$ is the surface gravity of the Schwarzschild black hole. Eq.\ (\ref{f(v)}) implicitly defines the spacetime trajectory of the accelerated mirror. Again by construction, our derived expression for $f(v)$ reduces to the Schwarzschild, Kerr and RN mirror trajectories in the relevant limits. A plot of these trajectories in a Penrose conformal diagram is shown in Fig.\ \ref{fig:penrose}. 

The rapidity as a function of advanced time, is given by $-2\eta(v) = \ln f'(v)$ where the prime denotes a derivative with respect to the argument. We obtain,
\begin{align}
    \eta(v) &= - \frac{1}{2} \ln \bigg| 1 - \frac{1}{\kappa_+v} + \frac{1}{\kappa_- (v - 4\lambda)} \bigg| .
\end{align}
Notably, the mirror approaches the speed of light as it approaches the horizon, $v\to 0$. The proper acceleration, $\alpha(v) = e^{\eta(v)} \eta'(v)$, can also be straightforwardly derived. For late-times, the acceleration, to leading order in $v$, is 
\begin{align}
    \alpha(v) &= - \frac{\kappa_+}{\sqrt{-4\kappa_+v}} + \mathcal{O}(v),  
\end{align}
which diverges (asymptotic infinite acceleration), and is dependent on $\kappa_+$. In the early-time limit, $v\to -\infty$, the proper acceleration vanishes.

%%%%%%%%%%%%%%%%%%%%%%%%%% 
\section{Energy Flux and Particle Spectrum} \label{sec:particles}
Having analysed the dynamics of the (1+1)-dimensional trajectory of the Kerr-Newman analog mirror, we now consider the properties of outgoing particle and energy fluxes induced by its motion. In the (1+1)-dimensional flat spacetime model, our analysis is restricted to the character of the analog Hawking radiation emitted by the mirror, neglecting scattering effects \cite{Fabbri}. In particular, our model does not describe the backscattering of wavepacket field modes into the black hole, nor superradiance, whereby incoming field modes are amplified due to the rotation of the black hole. Backscattering effects, manifesting through grey-body factors, are most prominent at low frequencies and negligible at high frequencies.  We are primarily interested in the non-scattering origin of pure Hawking radiation associated with freely propagating field modes (such as in near-horizon approximations).  Meanwhile, superradiance is a higher-dimensional effect which is mainly associated with the amplitude of the field modes and not the frequency as in the pure Hawking effect \cite{Fabbri}. Our analysis then focuses on the properties of the analog s-wave Hawking radiation, and the new insights which can be obtained in this simplified model.  

The radiated energy flux, $F(v)$, can be calculated from the quantum stress-energy tensor using the simple expression \cite{Good:2016atu}, 
\begin{align}
    F(v) &= \frac{1}{24\pi} \big\{f(v) ,v \big\} f'(v)^{-2},\label{stress}
\end{align}
where the Schwarzian brackets are defined as
\begin{align}
\big\{f(v), v \big\} &= \frac{f'''}{f'} - \frac{3}{2} \left( \frac{f''}{f'} \right)^2,
\end{align}
To leading order in $v$, near $v\to 0^-$, this yields
\begin{align}
    F(v) &= \frac{\kappa_+^2}{48\pi} + \mathcal{O}(v^2), \label{constantflux}
\end{align}
indicating uniform energy flux at late-times consistent with a temperature $T=\kappa_+/(2\pi)$. Eq.~(\ref{constantflux}) is identical to that found in \cite{Good:2020fjz} for the Kerr analog mirror, but includes a contribution from the charge within the surface gravity term. 

Next, we consider the particle spectrum of the outgoing modes. This can be derived from the Bogoliubov coefficients, 
%\begin{align}\label{16}
%    \beta_{\omega\omega' } &= - \frac{1}{4\pi \sqrt{\omega\omega'}} \int_{-\infty }^{v_H} \D v \:e^{-i\omega' v-i\omega f(v) } \big( \omega f'(v) - \omega' \big) 
%\end{align}
\begin{align}
    \beta_{\omega\omega' } &= \frac{1}{2\pi} \sqrt{\frac{\omega'}{\omega}} \int_{-\infty }^{v_H} \D v \:e^{-i\omega'v - i\omega f(v)},
\end{align}
where $\omega,\omega'$ are the frequencies of the outgoing and incoming modes respectively. This is a simplified form of the integral in e.g. \cite{Good:2016atu} where integration by parts neglects non-contributing surface terms.  The particle spectrum can be obtained by taking the modulus square, $N_{\omega\omega'}^\text{KN} = | \beta_{\omega\omega'} |^2$ which yields
\begin{align}\label{KNparticle}
    N_{\omega\omega'}^\text{KN} &= \frac{\omega'}{2\pi \kappa_+ \omega_+^2} \frac{e^{-\pi\omega /\kappa_-}}{e^{2\pi\omega/\kappa_+} - 1} | U |^2, 
\end{align}
where $\omega_+ = \omega + \omega'$ and
\begin{align}
    U \equiv U \left(  \frac{i\omega}{\kappa_-} ,  \frac{i\omega}{\kappa},  \frac{i\omega_+}{\bar{\kappa}} \right) ,
\end{align}
is a confluent hypergeometric Kummer function of the second kind. Here, $\bar{\kappa}^{-1} \equiv 2(r_+-r_{-}) = 4\lambda = 4\sqrt{M^2-a^2-Q^2}$. Eq.~(\ref{KNparticle}) is our first new result. 
\begin{figure}[h]
    \centering
    \includegraphics[width=0.95\linewidth]{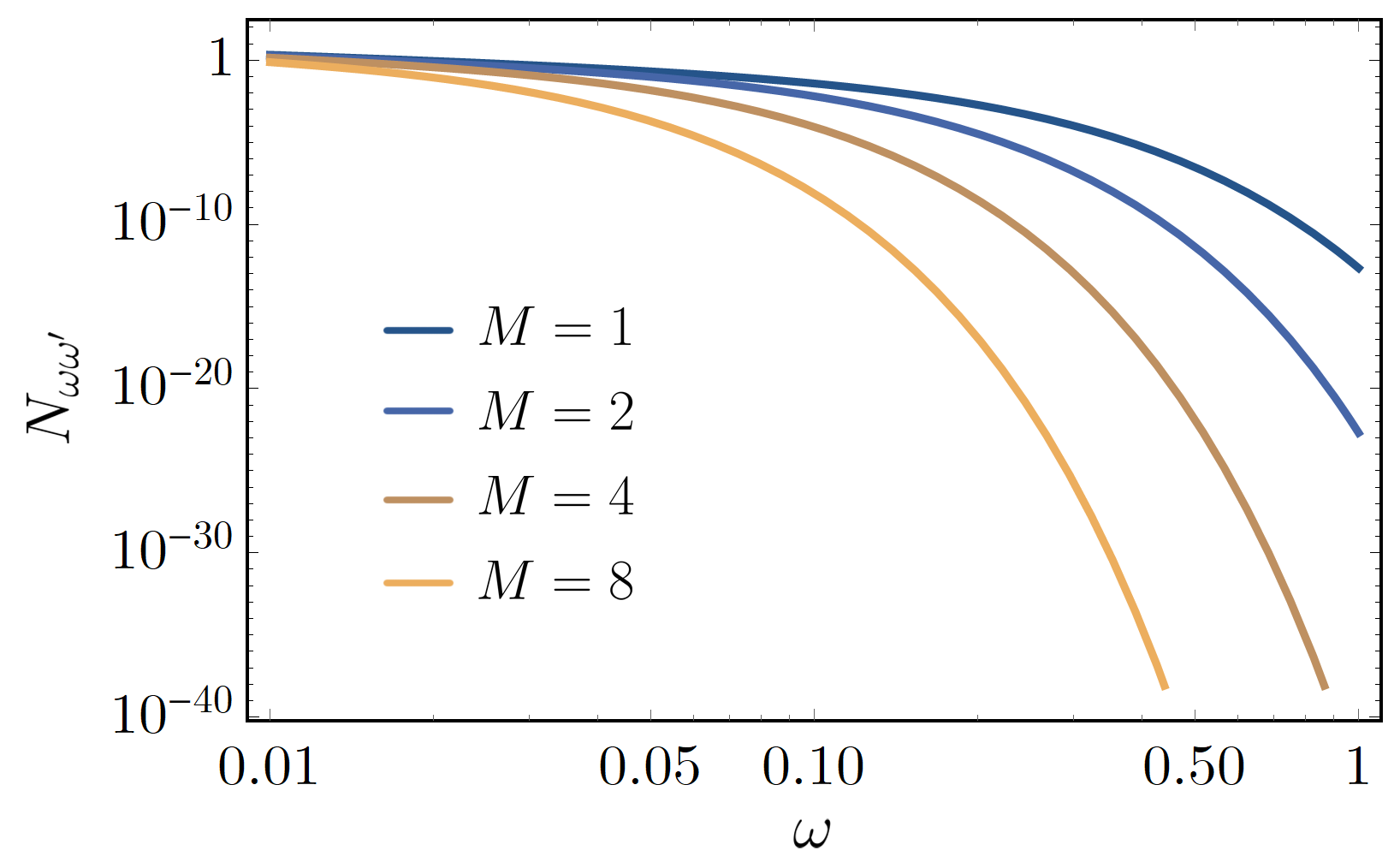}
    \caption{Mode-mode particle spectrum, $N_{\omega\omega'}$, plotted as a function of $\omega$ for different values of the black hole mass, $M$. We have used $a = Q = 0.5$. }
    \label{fig:Nwwspectrum}
\end{figure}
We associate the early-time regime with the limit $\omega \sim \omega'$, since the mirror is static in the asymptotic past and so the incoming and outgoing modes have approximately equal frequency. From Eq.\ (\ref{KNparticle}), the early-time spectrum is not thermal, which accords with known results like those derived using the Parikh-Wilczek tunnelling framework \cite{Parikh:1999mf,Zhang:2005uh,Jiang:2005ba}. 

In Fig.\ \ref{fig:Nwwspectrum}, we have plotted the mode-mode particle spectrum as a function of the outgoing frequency $\omega$, for fixed $\omega'$. Graphically, we find that the late-time spectrum $(\omega' \gg \omega)$ approaches an exactly thermal distribution in an analogous manner to the late-time Schwarzschild spectrum \cite{Good:2016oey}. This regime, as identified by Hawking \cite{Hawking:1974sw}, considers how the late-time incoming plane wave modes become significantly Doppler-shifted due to reflection off the asymptotically null mirror trajectory. Hence, the main contribution to the Bogoliubov coefficients comes from these high-frequency modes. We also prove this late-time behaviour analytically in Sec.\ \ref{sec:late-time}. 

\begin{figure}[h]
    \centering
    \includegraphics[width=\linewidth]{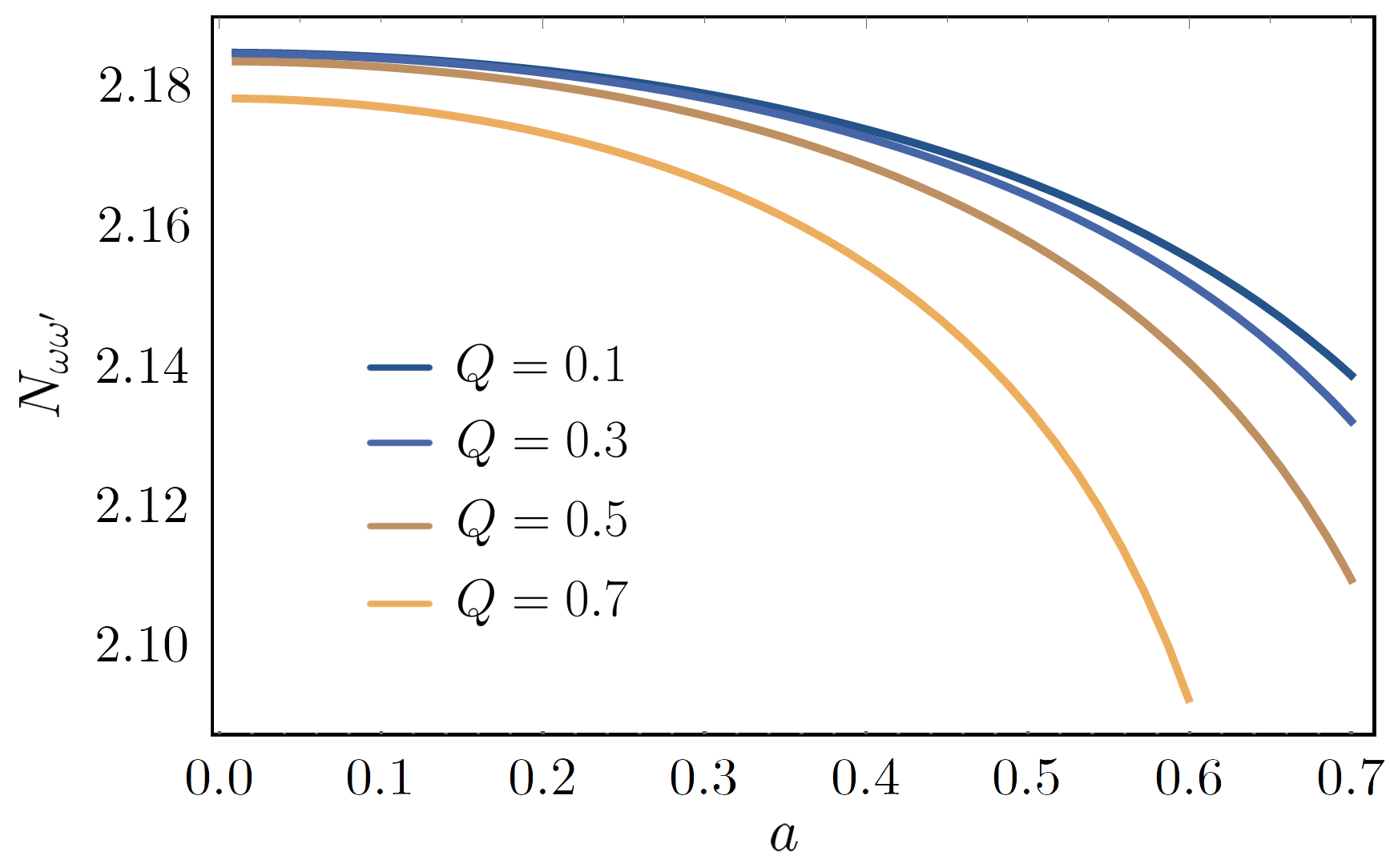}
    \caption{Mode-mode particle spectrum, $N_{\omega\omega'}$ as a function of angular momentum for late times, $\omega' = 1$, $\omega = 0.01$, with $M =1$ is fixed.}
    \label{fig:vsAQ}
\end{figure}

\begin{figure}[h]
    \centering
    \includegraphics[width=0.95\linewidth]{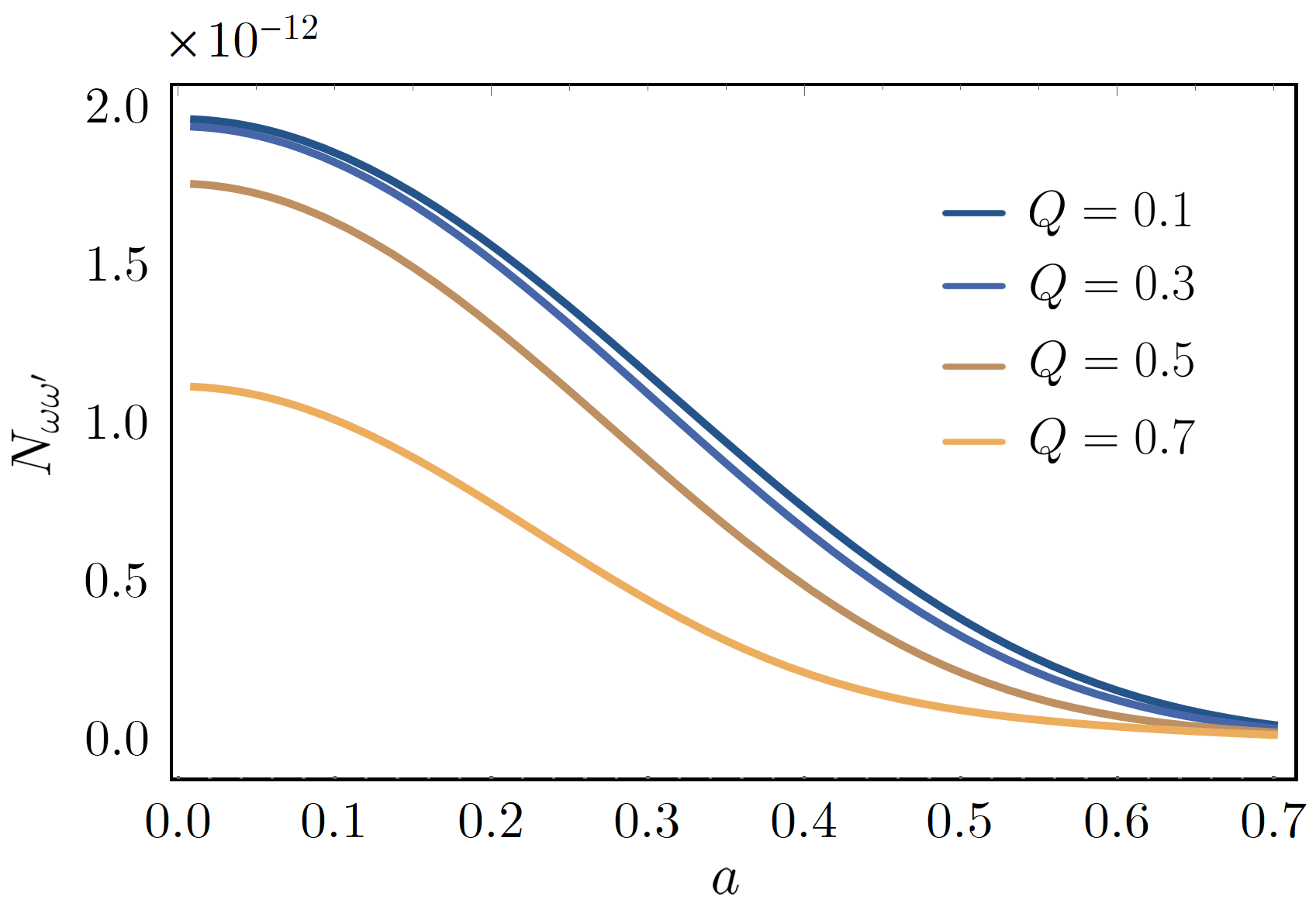}
    \caption{Mode-mode particle spectrum, $N_{\omega\omega'}$ as a function of angular momentum for early times, $\omega' = \omega =  1$, with $M =1$ is fixed.}
    \label{fig:vsAQearly}
\end{figure}

Figures \ref{fig:vsAQ} and \ref{fig:vsAQearly} plot the particle number as a function of the angular momentum, $a$, for the late- ($\omega'\gg \omega)$ and early-time ($\omega\sim\omega')$ regimes. In general, $N_{\omega\omega'}$ decays monotonically with increasing $a$, for all times. Likewise, the presence of charge inhibits particle production, most notably so for $Q \gtrsim 0.5$. 

\begin{figure}[h]
    \centering
    \includegraphics[width=0.8\linewidth]{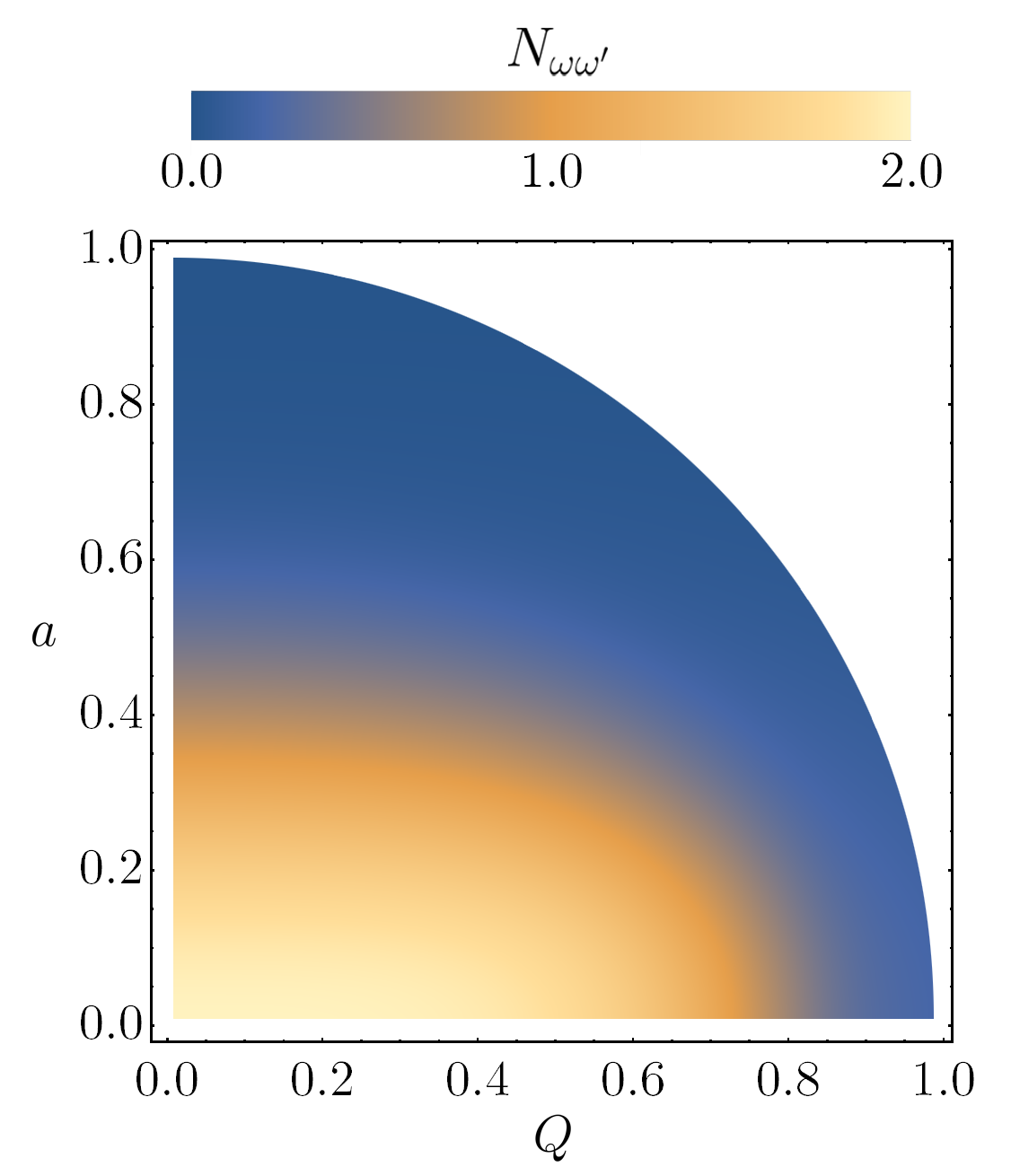}
    \caption{Mode-mode particle spectrum, $N_{\omega\omega'}$ as a function of $a,Q$ for early times, $\omega = \omega' = 1$. The white space corresponds to regions of the ($a,Q$) parameter space where $M^2 -a^2 - Q^2 < 0$, which is unphysical. The spectrum is normalized by $10^{-12}$ for illustration.}
    \label{fig:Nww}
\end{figure}

This behaviour is more clearly seen in the density plot of Fig.\ \ref{fig:Nww}, where $N_{\omega\omega'}$ is displayed as a function of both the charge, $Q$ and angular momentum, $a$ for early times. Notably, we find that the early-time particle spectrum is asymmetric with respect to the angular momentum and charge of the black hole. In particular, the rotation of the black hole inhibits particle production more strongly than an equivalent amount of charge. The faster decay of $N_{\omega\omega'}$ with respect to $a$ as compared to $Q$ may be due to the fact that for two equal mass Kerr-Newman black holes, the more rapidly rotating black hole is more energetic (see the Tolman-Landau-Lifshitz definition \cite{Virbhadra:1990vs}). Therefore, less energy is available to be carried away by the radiation. 

\subsection{Kerr Limit}
We now consider the limiting cases of Eq.\ (\ref{KNparticle}) which reduce to other well-known black hole solutions. In the limit $Q \to 0$, $N_{\omega\omega'}^\text{KN}$ reduces to the known result obtained for the analog Kerr system \cite{Good:2020fjz},
\begin{align}\label{Kerrlimit}
    \lim_{Q\to 0 } N_{\omega\omega'}^\text{KN} = N_{\omega\omega'}^\text{K} &= \frac{\omega'}{2\pi \kappa_+ \omega_+^2} \frac{e^{-\pi\omega/\kappa_-}}{e^{2\pi\omega/\kappa_+ }-1}| U^\text{K} |^2, 
\end{align}
where the inverse Kerr surface gravities and $U^\text{K}$ are
\be 
    \frac{1}{\kappa_\pm } = 2M \pm \frac{2M^2}{\sqrt{M^2 - a^2}}, \quad
%\end{align}
%and 
%\begin{align}
    U^\text{K} \equiv U \left( \frac{i\omega}{\kappa_-} ,  \frac{i\omega}{\kappa} , \frac{i\omega_+}{\bar{\kappa}} \right), 
\ee
%\end{align}
and are now written in terms of the relevant parameters in the Kerr limit. Here we have implicit uncharged Kerr quantities $\bar{\kappa}^{-1} = 2(r_+-r_{-}) = 4\sqrt{M^2-a^2}$. It can be straightforwardly verified that Eq.\ (\ref{Kerrlimit}) accords with that found in \cite{Good:2020fjz}. 

\subsection{Reissner–Nordstr\"om Limit}
Next, we take the $a\to 0$ (zero angular momentum) limit and find that 
\begin{align}\label{RNLimit}
    \lim_{a\to 0 } N_{\omega\omega'}^\text{KN} = N_{\omega\omega'}^\text{RN} &= \frac{\omega'}{2\pi \kappa_+ \omega_+^2} \frac{e^{-\pi\omega/\kappa_-}}{e^{2\pi\omega/\kappa_+ }-1}| U^\text{RN} |^2 
\end{align}
where
%\begin{align}
\be
    \frac{1}{\kappa_\pm} = 2M \pm \frac{2M^2 - Q^2}{\sqrt{M^2 - Q^2}},\quad  U^\text{RN} \equiv U\left(\frac{i\omega}{\kappa_{-}},\frac{i\omega}{\kappa}, \frac{i \omega_+}{\bar{\kappa}}\right),\ee
%\end{align}
are the inverse RN surface gravities, and Kummer function respectively.  Here $\bar{\kappa}^{-1} = 2(r_+ - r_{-})= 4\sqrt{M^2-Q^2}$.
We note here that Eq.\ (\ref{RNLimit}) is the same spectrum as found in \cite{good2020particle} but in more simple form due to integration of the Bogoliubov coefficient by parts and neglecting surface terms. 

\subsection{Schwarzschild \& Thermal Limits}\label{sec:late-time}
In the uncharged, non-rotating limit, the inner horizon vanishes and the particle spectrum reduces to 
\begin{align}
    \lim_{Q\to 0} \lim_{a\to 0 } N_{\omega\omega'}^\text{KN} = N_{\omega\omega'}^\text{S} &= \frac{\omega'}{2\pi \kappa \omega_+^2} \frac{1}{e^{2\pi\omega/\kappa}-1} ,
\end{align}
where $\kappa =1/4M$ is the usual surface gravity at the event horizon. This is exactly the Schwarzschild particle spectrum obtained in \cite{Good:2016oey,Good_2017Reflections,Anderson_2017,Good_2017BHII} where at late times $T = \kappa/(2\pi)$.

Finally, we consider the late-time limit for the outgoing particle spectrum, which we associate with $\omega'\gg \omega$ as identified by Hawking \cite{Hawking:1974sw}. As already mentioned, this regime can be associated with the late-time behaviour since the incoming plane wave modes are significantly Doppler-shifted due to the asymptotically null mirror trajectory. We have,
\begin{align}
    \lim_{\omega'\gg \omega } N_{\omega\omega'}^\text{KN} &= N_{\omega\omega'}^\text{CW} = \frac{1}{2\pi \kappa_+ \omega'} \frac{1}{e^{2\pi\omega/\kappa_+} -1}, 
\end{align}
where $\kappa_+$ is defined in Eq.\ (\ref{inversesurface}). Here, $N_{\omega\omega'}^\text{CW}$ is the eternal particle spectrum obtained for the Carlitz-Willey mirror trajectory \cite{carlitz1987reflections}, with temperature $T = \kappa_+/(2\pi)$. Of course, this is the same temperature as derived from the usual formula, 
\begin{align}
    \kappa_+ &= \frac{1}{2}\frac{\mathrm{d}}{\mathrm{d}r}f(r) \Big|_{r= r_+ } = \left( 2M + \frac{2M^2 - Q^2}{\sqrt{M^2 - a^2 - Q^2}} \right)^{-1} 
\end{align}
or for example, via the first law of black hole mechanics. This demonstrates that our newly derived mirror trajectories yield the correct temperature for the Kerr-Newman black hole, validating the accuracy of the correspondence between the two systems. 

\section{Extremal Kerr-Newman}\label{sec:extremal}
The extremal Kerr-Newman (EKN) limit represents the minimal possible mass which is compatible with the black hole's angular momentum and charge, and occurs for $M^2 = a^2 + Q^2$. Extremal black holes have been crucial in developing an understanding of the statistical origin of black hole entropy \cite{strominger1996microscopic}, making them relevant cases for studying quantum aspects of gravity. 

For the extremal Kerr-Newman case, one can obtain the relevant radial and time pieces of the metric in Eq.\ (\ref{KN1}), giving
\begin{align}
    f(r) &= 1 - \frac{2\sqrt{a^2 + Q^2}r- Q^2}{r^2 + a^2}
\end{align}
which yields the tortoise coordinate,
\begin{align}\label{tortoiseextremal}
    r^\star &= r - \frac{2a^2 + Q^2}{r-M} + 2M \ln \bigg| \frac{r-M}{2M} \bigg|
\end{align}
In Eq.\ (\ref{tortoiseextremal}), we have restored the extremal mass parameter $M$ where possible as shorthand for $M=+\sqrt{a^2+Q^2}$, keeping in mind the EKN system is characterized by only two free parameters, not three (KN). We have chosen an integration constant which yields the appropriate Schwarzschild limit in the non-rotating, uncharged case. 

Following the standard analysis by matching $r^\star = r$, solving for the radial coordinate of the origin for the black hole metric, and then applying the regularity condition of the modes, yields the following trajectory for the EKN mirror, 
%\begin{align}
%    f(v) &= v - \frac{4(2a^2 + Q^2)}{v} - 4M \ln \bigg| \frac{v}{4M} \bigg|.
%\end{align}
\begin{align}
    f(v) &= v - \frac{1}{\mathcal{A}^2 v} - \frac{1}{\kappa} \ln |\kappa v |.\label{f(v)EKN}
\end{align}
%\mike{This $M$ is the first and only $M$ in Eq.~(\ref{f(r)}).}\jf{Implicitly, we are saying that (in all these expressions) $M = \sqrt{a^2 + Q^2}$}. \jf{Or (two reduce the system to two parameters $a,M$):
%\begin{align}
%    f(v) &= v - \frac{4 ( a^2  + M^2)}{v} - 4M \ln \bigg| \frac{v}{4M} \bigg|
%\end{align}}
%\jf{Alternatively, one can write everything in terms of $a,Q$ to eliminate $M$, but the expressions are slightly messier since all the $M$'s are replaced by $M = \sqrt{a^2 + Q^2}$. What do you think? Alternatively, one could still write everything in terms of $a,M$ but make plots (like the density plot above) in terms of $a,Q$. }
where we have anticipated the parameter $\mathcal{A}$ as the asymptotic uniform acceleration, $\mathcal{A}^{-1} \equiv 2\sqrt{2a^2 + Q^2}=2\sqrt{a^2 +M^2}$.  Here $\kappa = 1/(4M)$ is the surface gravity of the Schwarzschild case and $M = +\sqrt{a^2 +Q^2}$.  One can straightforwardly verify that Eq.~(\ref{f(v)EKN}) reduces to the extremal RN solution \cite{good2020extreme} in the limit $a = 0$ and $M = Q$ and to the extremal Kerr solution \cite{Good:2020fjz} in the limit $Q=0$ and $M=a$.
% \begin{figure}[h]
%     \centering
%     \includegraphics[width=0.9\linewidth]{Penrose2.png}
%     \caption{Penrose conformal diagram of the analog mirror trajectories in the extremal limit. The trajectories, ranging from blue to orange, represent $a = 1/16, 1/8,1/4,1/2$ with $Q = 1/16$. \mike{This Fig appears to add little novelty from Fig 1.  Perhaps we dont need it or can find a way to make the curves drastically different at least via more extreme parametrization / colors / dashed lines. Thoughts?}}
%     \label{fig:my_label}
% \end{figure}
In particular, we have set the shell coordinate to be $v_0 = 2M$ with the horizon at $v_H = 0$. The rapidity is
%\begin{align}
%    \eta(v) &= - \frac{1}{2} \ln \bigg| \frac{4(2a^2 + Q^2) - v( v- 4M)}{v^2} \bigg|
%\end{align}
\begin{align}
    \eta(v) &= - \frac{1}{2} \ln \bigg| 1+ \frac{1}{\mathcal{A}^2 v^2} - \frac{1}{\kappa v} \bigg|,
\end{align}
%\jf{Or: %
%\begin{align}
%    \eta(v) &= - \frac{1}{2} \ln \bigg| \frac{4 (a^2 + M^2 )  - v(v- 4M)}{v^2} \bigg|
%\end{align}}
which describes a trajectory which is static in the asymptotic past and approaches the speed of light as $v\to 0^-$. In the asymptotic future, the trajectory approaches uniform acceleration, namely
\begin{align}\label{acceleration}
    \lim_{v\to 0 } \alpha(v) &= - \frac{1}{2\sqrt{2a^2 + Q^2}} \equiv -\mathcal{A}.
\end{align}
%\jf{Or: 
%\begin{align}
%    \lim_{v\to 0 } \alpha(v) &= - \frac{1}{2\sqrt{a^2 + M^2}} \equiv - \mathcal{A}
%\end{align}}
Thus, the EKN analog mirror advances toward asymptotic uniform acceleration. From the mirror trajectory, Eq.~(\ref{f(v)EKN}), we can determine the energy flux, Eq.~(\ref{stress}), given by 
% \iffalse 
% \begin{align}
%     F(v) &= \frac{v^3(M^2 v + 3Q^2v + a^2(6v - 8M) - M(4Q^2 + v^2)}{3\pi (4(2a^2 + Q^2 ) + v( v - 4M))^4}
% \end{align}
 
% \begin{align} \label{Fvextremal}
%     F(v) &= - \frac{v^3 ( a^2(4M - 3v) + M(v- 2M)^2)}{3\pi ( 4a^2 + (v-2M)^2 )^4}
% \end{align}
% \fi
%where we have expressed $F(v)$ in terms of $a,M$ for brevity.
% expressed in $\mathcal{A},M$, we obtain: \begin{align} \label{Fvextremal}
%     F(v) &= \frac{\mathcal{A}^6 v^3 \left(4 M \left(\mathcal{A}^2 v (M-v)-1\right)+3 v\right)}{12 \pi  \left(\mathcal{A}^2 v (v-4 M)+1\right)^4}
%     \end{align}
%   \mike{Benefits:  gives us that similar $\kappa^2/48\pi$ and also consistent with mirror parameters as $M$ no longer means mass in mirror system, consistent with 28\&29\&34, cost is that it less pretty
  \be
    F(v) = \frac{\kappa^2\mathcal{A}^6 v^3 \left(\mathcal{A}^2 v (1-4 \kappa  v)+4 \kappa  (3 \kappa  v-1)\right)}{48\pi \left(\mathcal{A}^2 v (\kappa  v-1)+\kappa \right)^4}\ee
\begin{figure}[h]
    \centering
    \includegraphics[width=\linewidth]{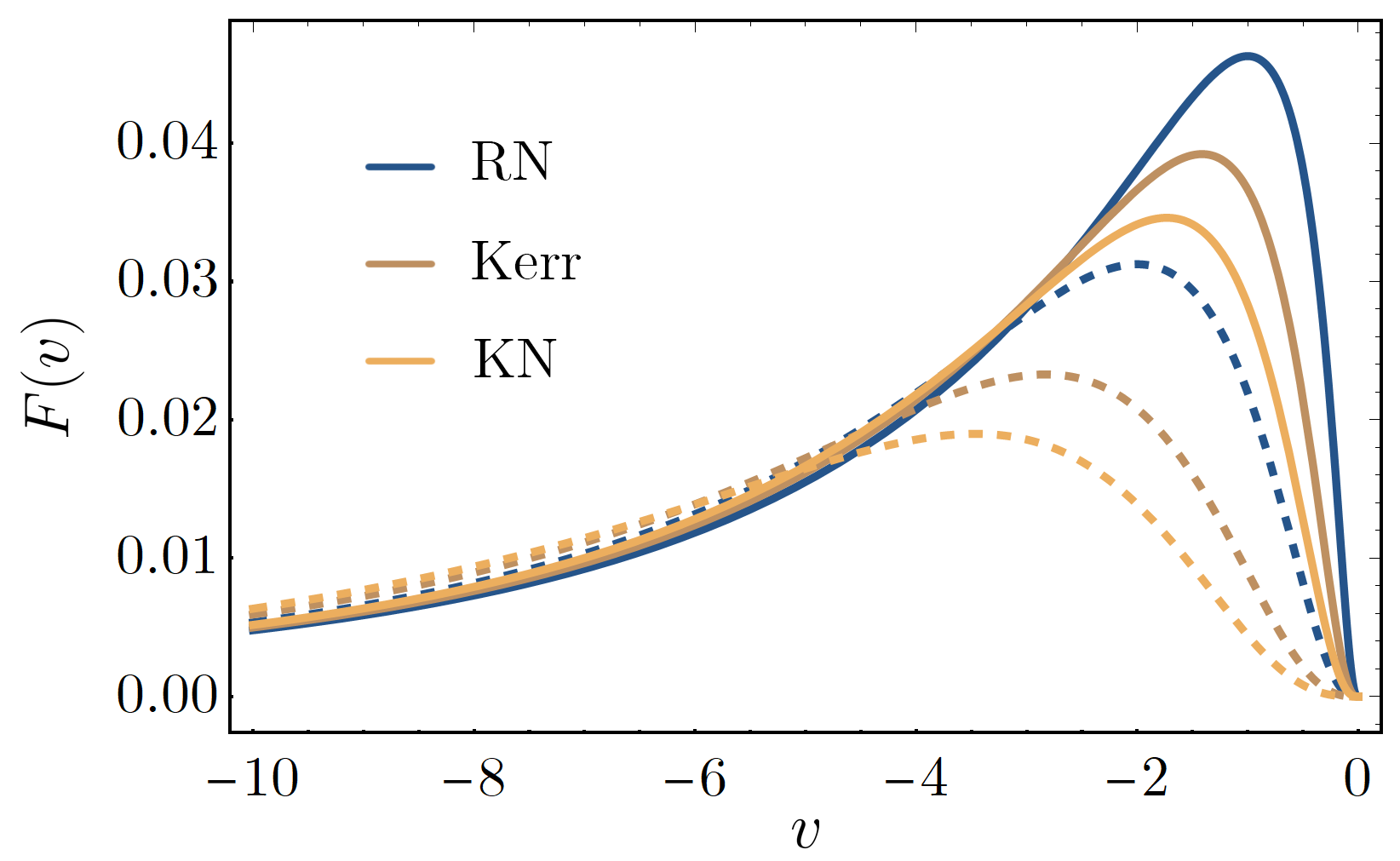}
    \caption{Energy flux, $F(v)$, normalised by $48\pi$, as a function of the advanced time coordinate $v$, for the (blue) RN with $Q = 0.5$ (solid) and $Q = 1.0$ (dashed), (brown) Kerr with $a = 0.5$ (solid) and $a = 1.0$ (dashed) and (orange) KN with $a = Q = 0.5$ (solid) and $a = Q = 1.0$ (dashed) trajectories.}
    \label{fig:energyflux}
\end{figure}
    
We find that $F(v)$ (plotted as a function of the advanced time, $v$ in Fig.\ \ref{fig:energyflux}) vanishes in the limit $v\to 0$, which is behavior in common with asymptotically inertial mirrors, even though the extremal mirror is asymptotically uniformly accelerating. Despite a terminating asymptotic energy flux, the model still produces an infinite total particle count.  Infinite total particle emission is not a property of asymptotic zero velocity (static) mirrors like the Walker-Davies model \cite{Walker_1982}, or the `Schwarzschild mirror with quantum purity' model \cite{Good:2019tnf,GoodMPLA, good2020schwarzschild}, which yield finite total particle count $N$.

\begin{figure}[h]
    \centering
    \includegraphics[width=0.975\linewidth]{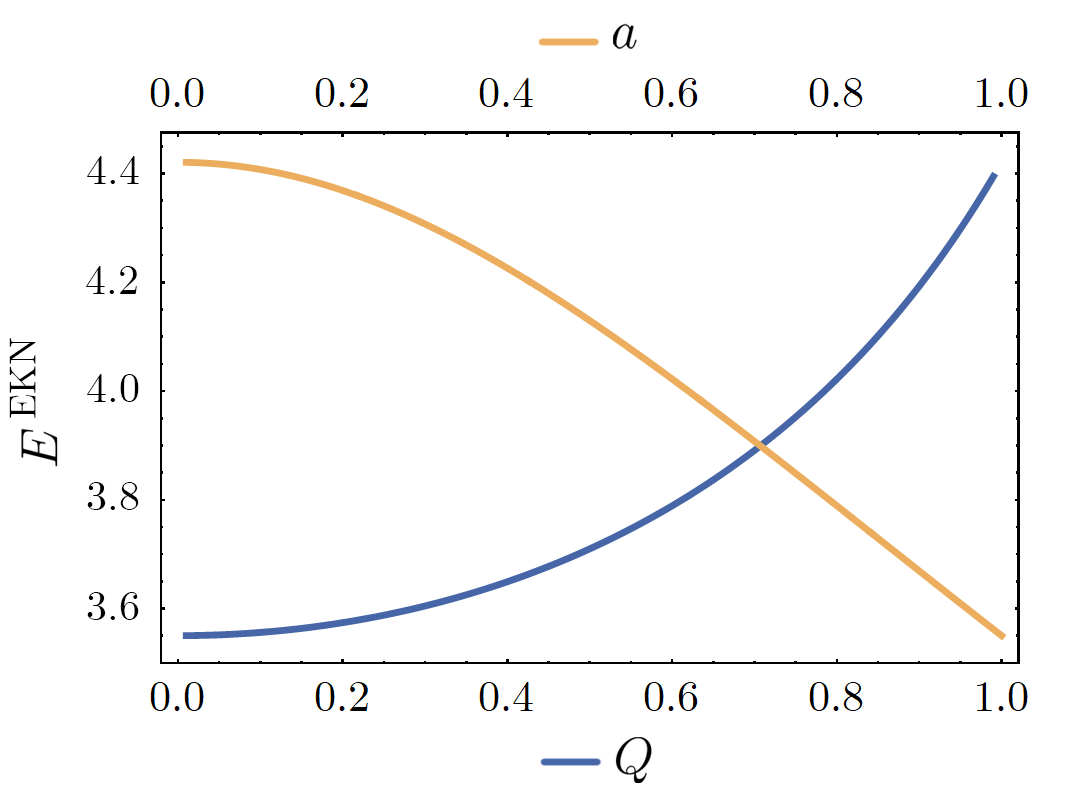}
    \caption{Total energy radiated by the analog extremal Kerr-Newman mirror, Eq.~(\ref{Totenergy}), with increasing $Q$ (blue) and $a$ (orange) and fixed $M = \sqrt{a^2 + Q^2}= 1$. The total energy, $E^\text{EKN}$, is scaled up by $10^3$. As $Q \to M$, the system reduces to the extremal RN case, while for $a \to 1$, we obtain the Kerr limit. For two equal mass black holes, the more charge, the more energy radiated.  Spin and charge are equal at $a=Q=M/\sqrt{2}$. }
    \label{fig:energy}
\end{figure}

The total energy is finite and analytic, found by integrating over the observer's retarded time $u$ at $\mathscr{I}^+_R$,
\begin{align}\label{stresstotenergyEKN}
    E &= \int_{-\infty }^{v_H = 0} F(v) \frac{\D f(v)}{\D v} \D v 
\end{align}
which yields
%\be 
%E = \frac{\left(3 a^2+M^2\right) \left(\frac{\pi }{2}-\cot ^{-1}\frac{a}{M}\right)-a M}{192\pi a^3}.\label{stresstotenergyEKN}\ee
%\mike{Maybe more simple:
\be
E =\frac{\kappa}{48\pi}\left[-\frac{1}{j^2}+\left(\frac{1}{j^3}+\frac{3}{j}\right)\tan ^{-1}j\right],\label{Totenergy}
\ee
where $j\equiv J/M^2 = a/M$ and $\kappa \equiv 1/(4M)$ is the surface gravity of the Schwarzschild black hole. We have plotted the total EKN energy, Eq.~(\ref{Totenergy}), as a function of $a,Q$ in Fig.\ \ref{fig:energy}. As before, $M^2 = a^2 + Q^2$ plays the role of the extremal mass parameter, so that only two free parameters enter the expression. Eq.~(\ref{Totenergy}) satisfies the appropriate limits in the extremal Kerr $(a\to 1)$ \cite{Good:2020fjz} and extremal RN \cite{good2020extreme} ($Q\to 1$) cases,
\be 
\lim_{Q\to 0^+} E = E^\text{EK},\qquad 
\lim_{a\to 0^+} E = E^\text{ERN} .
\ee
Furthermore, we observe from Fig.\ \ref{fig:energy} that the energy is bounded by the extremal Kerr and RN limits. Similar to the non-extremal case, there is asymmetry in the behavior of the total energy with respect to variations in $a$ and $Q$.
% \\\\
% \jf{Not sure about this, but from paper below (found via Scholarpedia article on KN metric) the classical energy of the exterior is 
% \begin{align}
%     E &= M - \frac{Q^2}{R} \left( \frac{a^2}{3R^2} + \frac{1}{2} \right)
% \end{align}
% where $R>r_+$ is outside the outer horizon. This is obtained via a definition of energy which demands the entire energy of a Schwarzschild black hole is
% confined to its interior only. This yields the following limits:
% \begin{align}
%     E(Q = 0) &= M \qquad \text{Kerr} \\
%     E(a = 0) &= M - \frac{Q^2}{2R} \qquad \text{RN}
% \end{align}
% The energy of rotating uncharged black holes is contained in its interior. The energy of charged, non-rotating black holes is shared by its exterior and interior. For rotation and charge, the first term in the brackets accounts for the magnetic field produced by the black hole. Numerically we can see that (for example) 
% \begin{align}
%     E(a = 0.1, Q = 0.5) &\sim 0.93 \\
%     E(a = 0.5, Q = 0.1) &\sim 0.99
% \end{align}
% The faster rotating black hole has more energy. Therefore less energy can be carried away by the radiation, compared to a case with the equivalent amount of charge. This may account for the faster decay of $N_{\omega\omega'}$ and $E$ with respect to $a$, as compared to $Q$.}\jf{The paper is here: \cite{Virbhadra:1990vs}}
Drawing an analogy with the black hole system, a finite energy ensures that the evaporation process stops, and that a naked singularity is impossible from the emission of neutral spin-0 particles. Like the ERN or EK, an infinite number of zero-frequency (soft) particles are radiated by the EKN, but only a finite amount of energy is carried away. 

The particle mode-mode spectrum is given by 
%\begin{align}
%    N_{\omega\omega'}^\text{KN} &= \frac{4e^{-4M\pi\omega} (2a^2 + Q^2) \omega' }{\pi^2 (\omega +\omega')} \bigg|K_{j} ( 4 \sqrt{\omega\omega_+ (2a^2 + Q^2)} \bigg|^2 
%\end{align}
%\jf{Or:
%\begin{align}
%    N_{\omega\omega'}^\text{KN} &= \frac{4e^{-4M\pi\omega}(a^2 + M^2)\omega'}{\pi^2(\omega + \omega')} \bigg| K_j \left( 4 \sqrt{\omega\omega_+ (a^2 + M^2)} \right) \bigg|^2  \\
%    &= \frac{e^{-4M\pi\omega} \omega'}{\pi^2\mathcal{A}^2\omega_+} \bigg| K_j \left( \frac{2}{\mathcal{A}} \sqrt{\omega\omega_+} \right) \bigg|^2 
%\end{align}}
\begin{align}
    N_{\omega\omega'}^\text{EKN} := |\beta_{\omega\omega'}^\text{EKN}|^2 &= \frac{e^{-\pi\omega/\kappa} \omega' }{\pi^2 \mathcal{A}^2 \omega_+} \bigg|K_{n}\left(\frac{2}{\mathcal{A}} \sqrt{\omega\omega_+}\right) \bigg|^2 
\end{align}
where $K_n(x)$ is the modified Bessel function of the second kind, with $n = 1- i\omega/\kappa$. Notably, this spectrum is explicitly non-thermal. The total energy carried by the particles, obtained via 
\begin{align}
    E &= \int_0^\infty\D \omega \int_0^\infty\D \omega'\, \omega \cdot N_{\omega\omega'}^\text{EKN}
\end{align}
is the same as Eq.~(\ref{Totenergy}) derived by the stress tensor radiation, Eq.~(\ref{stresstotenergyEKN}) confirming the signatures \cite{Good:2015nja} of energy flux in particle creation \cite{walker1985particle}. 

It should be clear that the extremal case has a {\it{particle}} spectrum which is never thermal and is not a limit of a thermal spectrum, still in agreement with \cite{Balbinot:2007kr} which found the smoothly connected limit of the stress energy tensor. The dynamics of the extremal analog mirror is entirely different to those found for the non-extremal case.  The extremal cases have late time constant acceleration, while the non-extremal cases have infinite late time acceleration. Apart from this drastic difference, our findings highlight the peculiarity of extremal black holes, separate from their non-extremal counterparts. Intuitively, the reason for this is a result of the characteristic of extremal black holes, namely a vanishing surface gravity, which yields an undefined temperature.

\section{Conclusions}\label{sec:conc} 
In this paper, we have derived a new accelerated boundary correspondence for the Kerr-Newman metric, describing rotating, charged black holes. In doing so, we have provided a simplified model of the Hawking radiation spectrum emitted from the black hole -- neglecting scattering effects. Our analytic result is expressed in terms of the hypergeometric Kummer function, and is found to be thermal in the late-time limit. 

The key novelties of our work are three-fold. Firstly, we have found a new application of the accelerated mirror model to mimic the particle spectrum of non-spherically symmetric spacetimes. This is a significant conceptual step that extends the scope of the well-known approach to modelling Hawking radiation. Secondly, we have derived a correspondence between the particle production of Kerr-Newman black holes to that obtained via the dynamical Casimir effect, for a new class of accelerated trajectories. Furthermore, our results for the spectrum in (1+1)-dimensions match those derived using the (3+1)-dimensional theory, in both the predicted temperature and for the Hawking radiation particle spectrum. Finally, we have derived new results for the particle and energy spectrum of the \textit{extremal} Kerr-Newman system, demonstrating for example that for equal mass black holes, larger amounts of charge lead to a higher total energy emission.

As illustrated, the approach applied here is highly accessible and can be easily extended to more complex spacetimes, such as black holes embedded within asymptotically de Sitter or anti-de Sitter spacetimes, or even regular Hayward and Bardeen black holes. We envision that new physics will continue to be discovered through this simple, but highly versatile model.

%%%%%%%%%%%%%%%%%%%%% 
\acknowledgments 

J.F.\ acknowledges support from the Australian Research Council Centre of Excellence for Quantum Computation and Communication Technology (Project
No.\ CE170100012). Funding from state-targeted program `Center of Excellence for Fundamental and Applied Physics' (BR05236454) by the Ministry of Education and Science of the Republic of Kazakhstan is acknowledged. M.G.\ is also funded by the FY2018-SGP-1-STMM Faculty Development Competitive Research Grant No. 090118FD5350 at Nazarbayev University.

\bibliography{main}

\end{document}